\documentclass{PoS}

\title{Optical variability modelling of newly identified blazar candidates behind Magellanic Clouds}

\ShortTitle{Optical variability modelling of blazar candidates behind MCs}

\author{\speaker{Natalia \.{Z}ywucka}$^{1,2}$, Mariusz Tarnopolski$^{2}$, Markus B\"{o}ttcher$^{1}$, {\L}ukasz Stawarz$^{2}$, Volodymyr Marchenko$^{2}$ \\
	    $^{1}$Centre of Space Research, North-West University, Potchefstroom, South Africa\\
        $^{2}$Astronomical Observatory, Jagiellonian University, ul. Orla 171, 30-244 Krak\'{o}w, Poland\\
        E-mail: \email{n.zywucka@oa.uj.edu.pl}}

\abstract{We present results of a variability study in the optical band of 44 newly identified blazar candidates behind the Magellanic Clouds. Our sample contains 27 flat spectrum radio quasars (FSRQs) and 17 BL Lacertae objects (BL Lacs). However, only nine of them are considered as secure blazar candidates, while the classification of the remaining 35 objects is still uncertain. All studied blazar candidates possess infrequently sampled optical light curves (LCs) in I filter provided by the Optical Gravitational Lensing Experiment group. The LCs were analysed with the Lomb-Scargle periodogram, the Hurst exponent $H$, and the $\mathcal{A}-\mathcal{T}$ plane, to look for blazar-like characteristic features and to study the long-term behaviour of the optical fluxes.

The power law (PL) indices of the Lomb-Scargle power spectral density (PSD) of the FSRQ blazar candidates mostly lie in the range $(1,2)$. In case of the BL Lacs they are located in the range $(1,1.8)$. The PL PSD is indicative of a self-affine stochastic process characterised by $H$, underlying the observed variability. We find that the majority of analysed objects have $H\leq 0.5$, indicating short-term memory, whereas four BL Lacs and two FSRQs have $H>0.5$, implying long-term memory. 41 blazar candidates are located in the $\mathcal{A}-\mathcal{T}$ plane in the region available to PL plus Poisson noise processes. Interestingly, one FSRQ is located marginally below this region, while two FSRQs lie above the line $\mathcal{T}=2/3$, i.e. they are even more noisy than white noise. The BL Lac candidates are characterised by higher $\mathcal{A}$ values than FSRQs, i.e. $0.71\pm 0.06$ and $0.29\pm 0.05$, respectively.
}

\FullConference{High Energy Phenomena in Relativistic Outflows VII - HEPRO VII\\
		9-12 July 2019\\
		Facultat de Física, Universitat de Barcelona, Spain}

\begin{document}

\section{Introduction}
Blazars constitute a sub-class of active galactic nuclei (AGNs) pointing their relativistic jets along the line of sight to an observer \cite{Ange80}. Based on the characteristic features visible in their optical spectra, blazars are commonly divided into two groups: flat spectrum radio quasars (FSRQs), having prominent emission lines with the equivalent width of $>$5 \r{A}, and BL Lacertae objects (BL Lacs) with featureless continua or weak emission lines only \cite{Urry95}. Blazars are characterised by non-thermal broad-band emission from radio up to $\gamma$-rays, high and variable polarization with the radio polarization degree usually defined at 1.4 GHz, PD$_{r,1.4}>$1\% \cite{Iler97}, flat radio spectra with the spectral index $\alpha_r<0.5$, and steep infrared to optical spectra, i.e. $0.5 \leq \alpha_o \leq 1.5$ \cite{Falo14}. Blazars show also rapid flux variability at all frequencies on different time scales from decades down to minutes.

In our previous work \cite{Zywu18}, we identified a sample of 44 blazar candidates, including 27 FSRQs and 17 BL Lacs. All objects in the sample were selected based on their radio, mid-infrared, and optical properties. First, we cross-matched radio positions from different catalogues with optical coordinates from the Magellanic Quasar Survey \cite{Kozl13}. Subsequently, we estimated and verified some characteristic properties of blazars, i.e. radio and mid-infrared indices, radio-loudness parameter, and fractional linear polarization. All blazar candidates in our sample are distant objects with redshifts from 0.29 to 3.32, optically faint with the I band magnitude between 17.66 and 21.27, and radio-loud with the radio-loudness parameter in the range $12-4450$ in the case of the FSRQ blazar candidates, and $171-7020$ for the BL Lac candidates. Based on archival data, we collected the radio polarimetry parameters for nine objects (six FSRQs and three BL Lacs). All are strongly polarised, with the average radio polarisation degree at 4.8 GHz of PD$_{\mathrm{r}, 4.8}\sim6.8\%$. Considering all the aforementioned parameters, i.e. $\alpha_r$, $\alpha_o$, $R$, and PD$_{\mathrm{r}, 4.8}$, these nine objects can be considered as secure blazar candidates. Here, we extend the analysis of our blazar candidates with modelling of optical LCs provided by the Optical Gravitational Lensing Experiment (OGLE) group. We investigate them to determine variability-based classification of the blazar candidates and to analyse long-term behaviour.

\section{The sample}

All objects included in our sample were selected from the well-monitored OGLE-III phase of the OGLE observations \cite{Udal08a,Udal08b}, which were conducted by the 1.3 m Warsaw telescope located at the Las Campanas Observatory in Chile in the I filter. The majority of objects were also monitored in the OGLE-IV phase \cite{Udal15}, while a few of them possess data from the OGLE-II phase \cite{Udal97} as well. In total, the blazar candidates with merged OGLE-II, OGLE-III, and OGLE-IV data have $\sim$17 years of LC coverage, those with OGLE-III and OGLE-IV stretch over $\sim$12 years, and with only OGLE-III data cover $\sim$7 years. 

\begin{figure*}
\includegraphics[width=\textwidth]{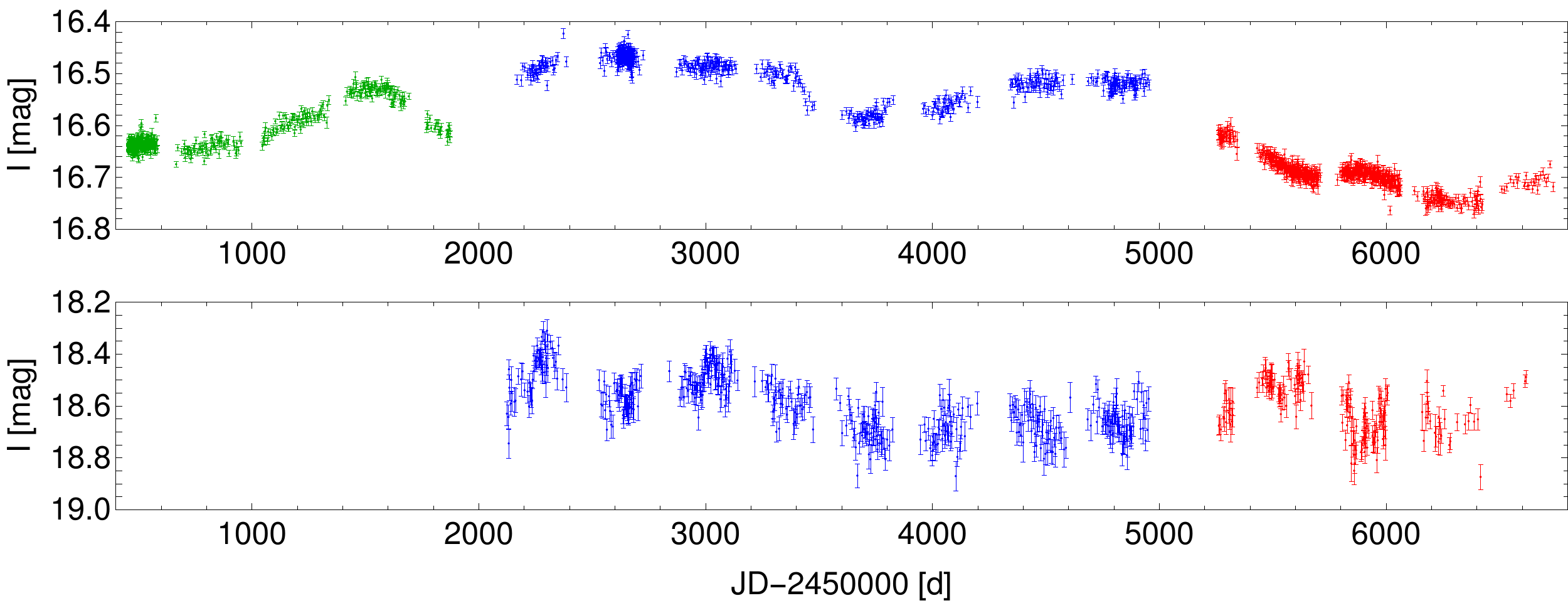}
\caption{Example of analysed LCs: J0532-6931 FSRQ candidate (top panel) and J0518-6755 BL Lac candidate (bottom panel). The OGLE-II data are shown with green colour, OGLE-III with blue colour, and OGLE-IV with red colour.}
\label{plot_LCs}
\end{figure*}

After visual inspection of the obtained photometric data we erased points with uncertainties $>$10\% in magnitude, manually removed outliers from all LCs, and excluded the OGLE-II data of one of the BL Lac candidates due to the high noise level. Rejection of points with uncertainties $>$10\% means that we lose only  $\sim$1\% of data, which does not significantly affect the results. An example of analysed LCs is shown in Fig.~\ref{plot_LCs}. All LCs are sampled irregularly with short, medium, and long time intervals between observations. Most of the objects were observed with a time step $\Delta t\approx 1\,{\rm d}$, with gaps lasting up to a few days. These short gaps were caused by bad weather conditions on the site. The medium time intervals are breaks in observations within the same OGLE phase, lasting between 3 and 5 months. During these times the MCs were too low to perform observations. Finally, after the OGLE-III phase, a technical upgrade of the telescope was performed, which resulted in a break (i.e. long time intervals) between the OGLE-III and OGLE-IV phases, lasting from 10 to 15 months.
 
\section{Methodology}

\subsection{Lomb-Scargle periodogram}

The Lomb-Scargle periodogram (LSP) \cite{Scar82,Vand18} is one of the methods to generate a power spectral density (PSD) for unevenly sampled time series. For an LC with $N$ observations $x_k$ at times $t_k$, it is computed as
\begin{equation}
P_{LS}(\omega) = \frac{1}{2\sigma^2} \left[ \frac{\left(\sum\limits_{k=1}^N (x_k-\bar{x}) \cos[\omega(t_k-\tau)]\right)^2}{\sum\limits_{k=1}^N \cos^2[\omega(t_k-\tau)]} \right. + \left. \frac{\left(\sum\limits_{k=1}^N (x_k-\bar{x}) \sin[\omega(t_k-\tau)]\right)^2}{\sum\limits_{k=1}^N \sin^2[\omega(t_k-\tau)]}  \right],
\end{equation}
where $\omega = 2\pi f$ is the angular frequency, $\tau\equiv\tau(\omega)$ is defined as
\begin{equation}
\tau(\omega)=\frac{1}{2\omega} \arctan \left[ \frac{\sum\limits_{k=1}^N\sin(2\omega t_k)}{\sum\limits_{k=1}^N\cos(2\omega t_k)} \right],
\end{equation}
$\bar{x}$ and $\sigma^2$ are the sample mean and variance, respectively. The following models are fitted to the LSP of each object:
\begin{enumerate}
    \item model A: a power law (PL) plus Poisson noise, estimated as $\frac{1}{2\sigma^2} \sum\limits_{k=1}^N \Delta x_k^2$, is given by
    \begin{equation}
        P(f) = \frac{P_{\rm norm}}{f^\beta}+C,
        \label{eqA}
    \end{equation}
where $\beta$ is the PSD exponent.
    \item model B: a smoothly broken PL (SBPL) plus Poisson noise \cite{Mcha2004}:
    \begin{equation}
        P(f) = \frac{P_{\rm norm}f^{-\beta_1}}{1 + \left( \frac{f}{f_{\rm break}} \right)^{\beta_2-\beta_1}} + C,
    \end{equation}
\end{enumerate}
where $f_{\rm break}$ is the break frequency and $\beta_1,\,\beta_2$ are the low and high frequency indices, respectively. 

Fits of different models are compared using the small sample Akaike Information Criterion:
\begin{equation}
AIC_c=2p-2\mathcal{L}+\frac{2(p+1)(p+2)}{N-p-2}
\label{}
\end{equation}
where $p$ is the number of parameters, $N$ is the number of fitted points, $\mathcal{L} = -\frac{1}{2}N\ln\frac{RSS}{N}$ is the log-likelihood, and $RSS$ is the residual sum of squares \cite{Akai74}. The relative goodness of a fit is assessed via the difference, $\Delta_i=AIC_{c,i}-AIC_{c,\rm min}$, between the $AIC_c$ of the $i$-th model and the one with the minimal value, $AIC_{c,\rm min}$. If $\Delta_i<2$, then both models are equally good. In such a case we choose model A as the simpler description. 

\subsection{Hurst exponent}

The Hurst exponent $H$ \cite{Hurs51} measures the statistical self similarity of a time series $x(t)$; $x(t)$ is self similar if it satisfies
\begin{equation}
x(t)\stackrel{\textbf{\textrm{.}}}{=}\lambda^{-H}x(\lambda t),
\end{equation}
where $\lambda>0$ and $\stackrel{\textbf{\textrm{.}}}{=}$ denotes equality in distribution. Self similarity is connected with long range dependence (memory) of a process via the autocorrelation function for lag $k$
\begin{equation}
\rho(k)=\frac{1}{2}\left[ (k+1)^{2H}-2k^{2H}+(k-1)^{2H} \right].
\end{equation}
The estimation of the $H$ values was performed with the wavelet lifting transform \cite{knight17}. 
\begin{enumerate}
\item $0<H<1$,
\item $H=0.5$ for an uncorrelated process,
\item $H>0.5$ for a persistent (long-term memory, correlated) process,
\item $H<0.5$ for an anti-persistent (short-term memory, anti-correlated) process.
\end{enumerate}

\subsection{The $\mathcal{A}-\mathcal{T}$ plane}
\label{meth_AT}

The Abbe value \cite{Neum41a,Tarn16d} is defined as
\begin{equation}
\mathcal{A}=\frac{\frac{1}{N-1}\sum\limits_{k=1}^{N-1}(x_{k+1}-x_k)^2}{\frac{2}{N}\sum\limits_{k=1}^N (x_k-\bar{x})^2}.
\end{equation}
It quantifies the smoothness of a time series by comparing the sum of the squared differences between two subsequent measurements with the standard deviation of the time series. Let $T$ denote the number of turning points in a time series and $\mathcal{T}=T/N$ be their frequency relative to the number of observations. The Gaussian noise is characterised by $\mathcal{T}=2/3$, while $\mathcal{T}>2/3$ is more noisy than white noise, and $\mathcal{T}<2/3$ is less noisy than Gaussian noise. 

\section{Results}

\begin{figure}[!h]
\includegraphics[width=\textwidth]{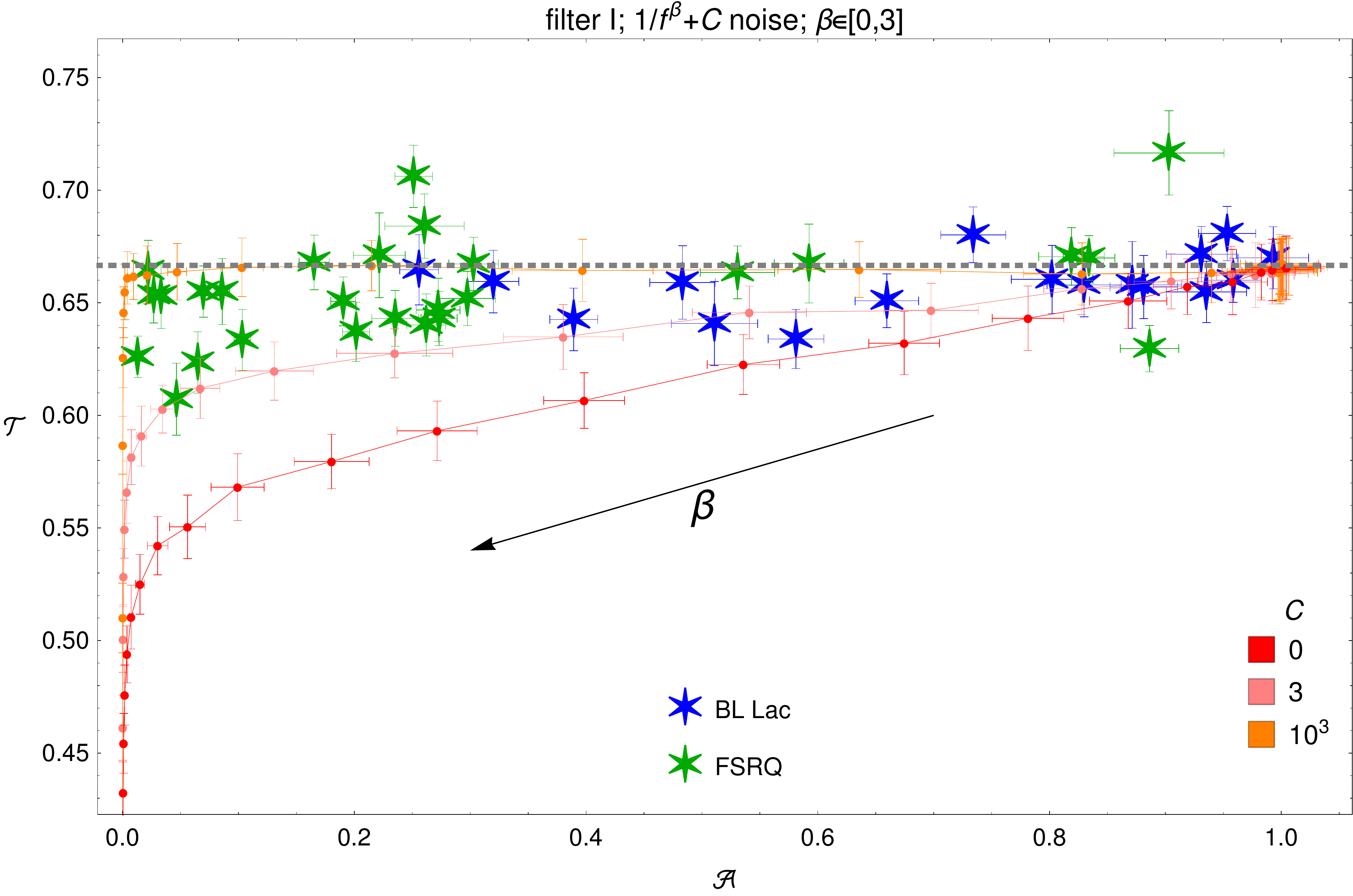}
\caption{Locations of our blazar candidates together with the PL plus Poisson noise PSDs in the $\mathcal{A}-\mathcal{T}$ plane. For each simulated PSD, 100 time series were generated and the displayed points are their mean locations. The horizontal gray dashed line denotes $\mathcal{T} = 2/3$ for a Gaussian process. The generic PL case, i.e. $C=0$, is shown with the red curve. The curves are raised and shortened when $C$ increases. The case $\beta=0$ is a white noise with ($\mathcal{A}, \mathcal{T}$) = (1, 2/3).}
\label{ThePlot}
\end{figure}

\begin{itemize}
\item \textbf{The LSP fitting}: model A is a better description of 10 FSRQs, while for 13 of them model B is preferred. For 13 out of 17 BL Lacs model A is in favour, however, for only two instances model B was preferred. The remaining 4 FSRQs and 2 BL Lacs are fitted well enough by both models. The indices derived with model A for the FSRQ blazar candidates mostly lie in the range $(1,2)$. BL Lacs are slightly flatter, spanning the range $(1,1.8)$; one object has a flat PSD, hence a pure PL was fitted. On the other hand, three BL Lacs have steeper PSDs, with $\beta \sim 3-4$. The derived exponents are listed in Table~\ref{AGNsamples}.
\item \textbf{The $H$ estimation}: for the majority of objects $H\leq 0.5$, indicating short-term memory. Four BL Lacs and two FSRQs yield $H>0.5$, implying long-term memory. A few objects are characterised by $H\approx 0.5$ (Table~\ref{AGNsamples}). 
\item \textbf{The $\mathcal{A}-\mathcal{T}$ plane}: among all 44 blazar candidates, 41 are located in the region occupied by PL plus $C$ (see Figure~\ref{ThePlot} and Table~\ref{AGNsamples}); one FSRQ is located marginally below, and two FSRQs are above the line $\mathcal{T}=2/3$. The BL Lac candidates are characterised by higher $\mathcal{A}$ values than FSRQs, i.e. $0.71\pm 0.06$ and $0.29\pm 0.05$, respectively.
\end{itemize}
 
\section{Conclusions}

\begin{enumerate}
\item 18 FSRQs in our sample yield PSDs consistent with model B. However, only 4 BL Lacs exhibit a detectable break in their PSDs. This might mean that the disk domination can manifest itself in the PSD via a break on the order of a few hundred days. Contrary, in case of BL Lacs, the lack of such a break might suggest jet domination. In this context, the four objects described by model B seem to be peculiar BL Lacs with a possibly significant radiative output coming from the accretion disk. 
\item The secure blazar candidates (see Table~\ref{AGNsamples}) have PSDs best described by model B, with $T_{\rm break}$ at $200-300\,{\rm days}$; one FSRQ and one BL Lac are consistent with model A.
\item In case of the blazar candidates best described by model B, the high frequency spectral index $\beta_2$ mostly lies in the range $(3,7)$. This can indicate a new class of AGNs in which the short term variability is effectively wiped out.
\item Two FSRQs and four BL Lacs were found to exhibit $H>0.5$, indicating long-term memory of the underlying governing process. This suggests that more complicated stochastic models need to be considered as a source of the observed variability.
\item The recently developed $\mathcal{A}-\mathcal{T}$ plane was used here in order to classify LCs. We identified two FSRQs located in a region not available for PL types of PSD, i.e. J0512$-$7105 and J0552$-$6850. The former exhibits a flat PSD, while the latter yields an SBPL with $T_{\rm break}>1000\,{\rm days}$.

\end{enumerate}

\newpage

\begin{table}[!h]
\caption{Parameters of the fits of newly identified FSRQ and BL Lac type blazar candidates.}
\label{AGNsamples}
\scriptsize
\begin{center}
\begin{tabular}{cccccccccc}
Object & $\beta$ & $\beta_1$ & $\beta_2$ & $T_{\rm break}$ & Best & $H$ & $\mathcal{A}$ & $\mathcal{T}$  \\
 & & & & [d] & model &  \\
 \tiny{(1)} & \tiny{(2)} & \tiny{(3)} & \tiny{(4)} & \tiny{(5)} & \tiny{(6)} & \tiny{(7)} & \tiny{(8)} & \tiny{(9)} \\\hline  
\multicolumn{8}{c}{FSRQ type blazar candidates}\\\hline
 J0054$-$7248 & $1.21\pm 0.13$ & --- & --- & --- & A & $0.42\pm 0.02$ & $0.83\pm 0.02$ & $0.670\pm 0.011$ \\
 J0114$-$7320$^{\star}$ & $1.45\pm 0.18$ & $0.37\pm 0.40$ & $4.53\pm 1.56$ & $338\pm 127$ & B & $0.06\pm 0.04$ & $0.033\pm 0.002$ & $0.653\pm 0.015$ \\
 J0120$-$7334$^{\star}$ & $1.86\pm 0.15$ & $1.28\pm 0.23$ & $5.00\pm 1.91$ & $229\pm 75$ & B & $0.04\pm 0.03$  & $0.027\pm 0.002$ & $0.653\pm 0.012$ \\
 J0122$-$7152 & $1.45\pm 0.17$ & $1.01\pm 0.27$ & $5.65\pm 4.35$ & $155\pm 61$ & A, B & $0.24\pm 0.04$  & $0.24\pm 0.02$ & $0.643\pm 0.012$ \\
 J0442$-$6818$^{\star}$ & $1.75\pm 0.27$ & $0.83\pm 0.31$ & $8.93\pm 5.20$ & $241\pm 51$ & B & $0.04\pm 0.03$  & $0.022\pm 0.002$ & $0.664\pm 0.014$ \\
 J0445$-$6859 & $1.30\pm 0.29$ & --- & --- & --- & A & $0.31\pm 0.05$ & $0.59\pm 0.03$ & $0.667\pm 0.017$ \\
 J0446$-$6758 & $1.60\pm 0.18$ & --- & --- & --- & A & $0.45\pm 0.05$ & $0.30\pm 0.02$ & $0.652\pm 0.011$  \\
 J0455$-$6933 & $1.58\pm 0.28$ & $0.30\pm 0.36$ & $6.83\pm 3.38$ & $250\pm 58$ & B & $0.25\pm 0.05$ & $0.22\pm 0.02$ & $0.671\pm 0.019$ \\  J0459$-$6756 & $1.64\pm 0.27$ & $0.88\pm 0.44$ & $6.56\pm 5.47$ & $246\pm 103$ & A, B & $0.21\pm 0.03$ & $0.17\pm 0.01$ & $0.668\pm 0.012$  \\
 J0510$-$6941 & $1.63\pm 0.13$ & $0.96\pm 0.16$ & $5.75\pm 1.58$ & $218\pm 39$ & B & $0.04\pm 0.03$ & $0.20\pm 0.01$ & $0.637\pm 0.013$ \\
 J0512$-$7105 & $0.14\pm 0.06$ & --- & --- & --- & A & $0.63\pm 0.05$  & $0.90\pm 0.05$ & $0.717\pm 0.019$ \\
 J0512$-$6732$^{\star}$ & $1.00\pm 0.12$ & $0.64\pm 0.09$ & $6.84\pm 3.26$ & $67\pm 10$ & B & $0.85\pm 0.03$ & $0.26\pm 0.02$ & $0.640\pm 0.014$ \\
 J0515$-$6756 & $1.40\pm 0.24$ & --- & --- & --- & A & $0.38\pm 0.02$ & $0.82\pm 0.03$ & $0.670\pm 0.012$  \\
 J0517$-$6759 & $1.23\pm 0.21$ & $0.70\pm 0.30$ & $6.66\pm 6.04$ & $164\pm 56$ & A, B & $0.29\pm 0.04$ & $0.53\pm 0.03$ & $0.663\pm 0.013$\\
 J0527$-$7036 & $1.26\pm 0.13$ & $1.06\pm 0.22$ & $3.82\pm 3.69$ & $48\pm 34$ & A & $0.04\pm 0.03$ & $0.07\pm 0.01$ & $0.655\pm 0.013$\\
 J0528$-$6836 & $1.47\pm 0.15$ & --- & --- & --- & A & $0.26\pm 0.05$ & $0.33\pm 0.02$ & $0.656\pm 0.015$ \\
 J0532$-$6931 & $1.29\pm 0.14$ & $0.68\pm 0.21$ & $4.33\pm 1.76$ & $115\pm 45$ & B & $0.07\pm 0.03$ & $0.013\pm 0.001$ & $0.626\pm 0.009$  \\
 J0535$-$7037 & $1.11\pm 0.14$ & --- & --- & --- & A & $0.42\pm 0.03$ & $0.89\pm 0.03$ & $0.630\pm 0.011$ \\
 J0541$-$6800 & $1.56\pm 0.15$ & $0.71\pm 0.34$ & $3.35\pm 1.00$ & $319\pm 172$ & B & $0.08\pm 0.05$ & $0.30\pm 0.02$ & $0.667\pm 0.012$\\
 J0541$-$6815 & $1.92\pm 0.12$ & $1.45\pm 0.16$ & $5.87\pm 1.76$ & $177\pm 34$ & B & $0.21\pm 0.03$ & $0.27\pm 0.02$ & $0.643\pm 0.012$  \\
 J0547$-$7207 & $1.37\pm 0.21$ & $0.29\pm 0.37$ & $4.66\pm 2.00$ & $284\pm 106$ & B & $0.03\pm 0.02$ & $0.09\pm 0.01$ & $0.655\pm 0.014$\\
 J0551$-$6916$^{\star}$ & $1.46\pm 0.22$ & $0.75\pm 0.31$ & $7.36\pm 5.21$ & $225\pm 64$ & B & $0.06\pm 0.04$ & $0.046\pm 0.004$ & $0.607\pm 0.016$\\
 J0551$-$6843$^{\star}$ & $1.48\pm 0.17$ & --- & --- & --- & A & $0.11\pm 0.04$ & $0.065\pm 0.005$ & $0.623\pm 0.014$\\
 J0552$-$6850 & $1.62\pm 0.14$ & $-0.70\pm 0.76$ & $2.36\pm 0.28$ & $1201\pm 417$ & B & $0.22\pm 0.05$ & $0.25\pm 0.02$ & $0.706\pm 0.013$\\
 J0557$-$6944 & $1.57\pm 0.24$ & --- & --- & --- & A & $0.49\pm 0.05$ & $0.26\pm 0.03$ & $0.684\pm 0.014$\\
 J0559$-$6920 & $1.44\pm 0.19$ & $0.79\pm 0.33$ & $5.51\pm 3.57$ & $248\pm 93$ & A, B & $0.22\pm 0.03$ & $0.03\pm 0.02$ & $0.647\pm 0.014$\\
 J0602$-$6830 & $1.35\pm 0.13$ & $0.30\pm 0.69$ & $2.25\pm 0.68$ & $538\pm 579$ & B & $0.07\pm 0.04$ & $0.10\pm 0.01$ & $0.634\pm 0.014$\\\hline
\multicolumn{8}{c}{BL Lac type blazar candidates}\\\hline
 J0039$-$7356 & $1.61\pm 0.28$ & --- & --- & --- & A & $0.44\pm 0.03$ & $0.96\pm 0.02$ & $0.660\pm 0.010$\\
 J0111$-$7302$^{\star}$ & $1.76\pm 0.44$ & --- & --- & --- & A & $0.35\pm 0.04$ & $0.73\pm 0.03$ & $0.680\pm 0.012$\\
 J0123$-$7236 & $4.02\pm 1.23$ & --- & --- & --- & A & --- & $0.95\pm 0.03$ & $0.681\pm 0.012$\\
 J0439$-$6832 & $0.98\pm 0.22$ & --- & --- & --- & A & $0.60\pm 0.06$ & $0.83\pm 0.03$ & $0.658\pm 0.014$\\
 J0441$-$6945 & $1.20\pm 0.16$ & --- & --- & --- & A & $0.45\pm 0.03$ & $0.80\pm 0.04$ & $0.660\pm 0.014$\\
 J0444$-$6729 & $1.47\pm 0.21$ & $1.00\pm 0.48$ & $4.42\pm 4.20$ & $193\pm 133$ & A & $0.21\pm 0.05$ & $0.51\pm 0.04$ & $0.641\pm 0.019$\\
 J0446$-$6718 & $3.50 \pm 1.13$ & --- & --- & --- & A & $0.58\pm 0.05$ & $0.94\pm 0.03$ & $0.655\pm 0.015$\\
 J0453$-$6949 & $2.64\pm 0.66$ & --- & --- & --- & A & $0.48\pm 0.04$ & $0.93\pm 0.03$ & $0.672\pm 0.012$  \\
 J0457$-$6920 & $1.03\pm 0.18$ & --- & --- & --- & A & $0.26\pm 0.03$ & $0.66\pm 0.03$ & $0.651\pm 0.012$  \\
 J0501$-$6653$^{\star}$ & $1.44\pm 0.20$ & $0.98\pm 0.32$ & $6.95\pm 6.63$ & $217\pm 78$ & A, B & $0.29\pm 0.04$ & $0.39\pm 0.02$ & $0.643\pm 0.014$ \\
 J0516$-$6803 & $-0.04\pm 0.05$ & --- & --- & --- & A  & $0.62\pm 0.05$ & $0.99\pm 0.03$ & $0.670\pm 0.014$\\
 J0518$-$6755$^{\star}$ & $1.34\pm 0.15$ & $0.84\pm 0.33$ & $3.75\pm 2.28$ & $182\pm 118$ & A, B & $0.18\pm 0.03$ & $0.26\pm 0.02$ & $0.665\pm 0.015$\\
 J0521$-$6959 & $1.16\pm 0.24$ & --- & --- & --- & A & $0.23\pm 0.04$ & $0.48\pm 0.03$ & $0.659\pm 0.016$ \\
 J0522$-$7135 & $1.16\pm 0.39$ & --- & --- & --- & A & $0.39\pm 0.04$ & $0.88\pm 0.03$ & $0.657\pm 0.014$\\
 J0538$-$7225 & $1.09\pm 0.16$ & $0.42\pm 0.25$ & $5.17\pm 2.86$ & $183\pm 57$ & B & $0.28\pm 0.04$ & $0.32\pm 0.02$ & $0.659\pm 0.014$\\
 J0545$-$6846 & $0.99\pm 0.38$ & --- & --- & --- & A & $0.58\pm 0.05$ & $0.87\pm 0.04$ & $0.658\pm 0.019$ \\
 J0553$-$6845 & $1.34\pm 0.19$ & --- & --- & --- & A & $0.36\pm 0.04$ & $0.58\pm 0.02$ & $0.634\pm 0.013$\\\hline
\end{tabular}   
\end{center}
\begin{minipage}{1.0\textwidth}
$^{\star}$Strongly polarized sources at radio frequencies which are considered as secure blazar candidates by \cite{Zywu18}.\\Columns: (1) source designation; (2) PL index of model A; (3) low frequency index of model B; (4) high frequency index of model B; (5) break time scale of model B; (6) best model; (7) The Hurst exponent; (8) the Abbe value; (9) ratio of turning points.
\end{minipage}
\end{table}

\acknowledgments
The authors thank the OGLE group for the photometric data of the analysed sources. N\.{Z} work was supported by the Polish National Science Center (NCN) through the PRELUDIUM grant DEC-2014/15/N/ST9/05171. MT acknowledges support by the NCN through the OPUS grant No. 2017/25/B/ST9/01208. The work of MB is supported through the South African Research Chair Initiative of the National Research Foundation\footnote{Any opinion, finding and conclusion or recommendation expressed in this material is that of the authors and the NRF does not accept any liability in this regard.} and the Department of Science and Technology of South Africa, under SARChI Chair grant No. 64789. {\L}S and VM acknowledge support by the NCN through grant No. 2016/22/E/ST9/00061.

\end{document}